\documentclass{icrc29}

\usepackage{graphicx}

\usepackage{epsfig}

\begin{document}

\title{A new estimate of the Galactic interstellar radiation field between
0.1 $\mu$m and 1000 $\mu$m}

\author[T. A. Porter \& A. W. Strong] {T. A. Porter$^a$, A. W. Strong$^b$ \\
(a) Department of Physics and Astronomy, Louisiana State University, Baton Rouge LA70803, USA. \\
(b) Max-Planck Institut f\"ur Extraterrestrische Physik, Garching, Germany 
}

\presenter{T. A. Porter (tporter@lsu.edu), \
usa-porter-T-abs1-og21-poster}

\maketitle

\begin{abstract}
Cosmic-ray electrons and positrons propagating in the Galaxy produce
diffuse gamma-rays via the inverse Compton (IC) process. 
The low energy target photon populations with which the cosmic-rays interact 
during propagation are produced by stars, this stellar light being 
reprocessed by Galactic dust. 
Detailed modelling of the Galactic stellar distribution, dust
distribution, and treatment of the absorption and scattering of light is
therefore required to obtain accurate models for the low energy Galactic
photon distribution and 
spectrum.
Using a realistic Galactic stellar distribution model, and dust distribution, 
we calculate the diffuse radiation field from stars in the Galaxy (the
`optical' radiation field), including absorption and scattering. 
Using a dust heating code, we self-consistently calculate the infra-red 
radiation
field for the same dust model used for the optical calculation; both
transient and equilibrium heating are included. 
We present the calculated
radiation field spectra and distributions, and will use these to 
calculate the  
expected Galactic diffuse IC gamma-ray spectrum.
\end{abstract}

\section{Introduction}
The Galactic interstellar radiation field (ISRF) plays an important role
in the propagation (energy losses) and $\gamma$-ray emission of cosmic-ray 
electrons, via the inverse Compton effect.
For $\gamma$-ray calculations we require the spectrum of the ISRF in three 
dimensions over the whole Galaxy.
The most detailed calculation to date (Strong et al. 2000) has been widely 
used, but subsequent to this work a large amount of relevant new astronomical 
information on stellar populations, Galactic structure and interstellar dust 
has become available.
Here we give a status report on our efforts to compute a new ISRF, exploiting 
this new information, for use with upcoming $\gamma$-ray missions like GLAST.

\section{Modelling the Interstellar Radiation Field}
We use a model for the distribution of stars within the Galaxy
based on the statistical SKY model of Wainscoat et al. (1992). 
We have also included the improvements to the model given by 
Cohen (1993, 1994, 1995). 
The stellar model assumes a type classification, which includes 
87 stellar classes encompassing main sequence stars, AGB stars, T-Tauri 
stars, and planetary nebula.
For each stellar type, there is an associated local density, scale height, 
fraction of local density in each of several discrete spatial components 
(see later), and spectrum given in the Johnson-Cousins-Glass photometric 
system, plus filters at 12 $\mu$m and 25 $\mu$m.

We have made several modifications to the original model to better match
recent observations and theoretical advances in stellar spectral modelling :

\begin{enumerate}
\item{The original model included five discrete spatial components : bulge,
arms, disc, halo, and ring.
Recent analysis of 2MASS data (L\'{o}pez-Corredoira et al. 2001) 
indicates the presence of an offset bar, distinct from the bulge.
The bar is a thin cylindrical structure with the near side offset with
respect to the galactic centre (GC)-Sun line by $l \sim -20^\circ$ to 
$-30^\circ$.
We model the bar as a thin ellipsoid with parameters as described by 
Freudenreich (1998), and assume the bar stellar 
population is similar to the disc, with normalisation similar to the bulge.}

\item{Analysis of COBE data (Freudenreich 1998) 
and DENIS star counts (L\'{o}pez-Corredoira et al. 2002) 
reveals a cut-off within a few kpc from the GC, and no cut-off
out to a distance of at least 15 kpc; see also 
L\'{o}pez-Corredoira et al. (2004).
Ojha (2001) obtains from 2MASS data a disc scale-length of 
$\sim 2.8$ kpc.
Therefore, we model the disc with an exponential cut-off inside 3 kpc, 
and a smooth exponential distribution with scale-length 3 kpc for distances
greater than 3 kpc from the GC.}

\item{The original SKY model spectra were empirically derived from 
observations of stars corresponding to each respective stellar type, 
or, for AGB stars, from a stellar evolution model.
For normal stars, we have instead obtained spectra from the 
synthetic spectral library of Girardi et al. (2002).
For each stellar type, we obtained reference values for effective temperature, 
luminosity, and surface gravity from Allen (2000).
Then, a representative spectrum for each stellar type is obtained by 
interpolating over the Girardi grid.
If an appropriate spectrum cannot be found in the Girardi grid, the original
SKY model spectrum is used.}

\end{enumerate}

The light emitted by stars is absorbed and scattered by dust in the 
interstellar medium (ISM).
Absorbed light is re-emitted in the infra-red, while scattered light undergoes
further absorption and scattering.
To calculate the dust extinction (absorption and scattering), and 
diffuse infra-red emission, we require a model for dust in the ISM.
We include in our model graphite grains, polycyclic aromatic hydrocarbons 
(PAHs), and silicate grains, since some mixture including these, or similar 
molecules, reproduces the galactic extinction curve (Zubko et al. 2004).
We assume that grains in our model are spherical, and take the optical
absorption and scattering efficiencies for graphite, PAHs, 
and silicate grains from Li \& Draine (2001).
We adopt the grain model abundance and size distribution from
Weingartner \& Draine (2001) (their best fit Milky Way model).

Heating of dust grains in the ISM by the ambient optical radiation field 
produces the diffuse infra-red emission.
The dust absorption and re-emission process is dependent on the size of the
dust grains involved.
Grains with sizes less than $\sim 0.1$ $\mu$m (`small') 
are stochastically heated, and undergo temperature fluctuations. 
Grains with sizes larger than $\sim 0.1$ $\mu$m (`large') 
maintain a nearly constant temperature under heating.
For `small' grains, we calculate the transient heating process and 
re-emission 
using the `thermal continuous' approach of Draine \& Li (2001).
For `large' grains, we obtain the infra-red emission by balancing absorption 
with re-emission as described by Li \& Draine (2001).

Scattering by dust in the ISM produces the so-called diffuse galactic light
(DGL) that comprises about $10\% - 30\%$ of the total optical radiation 
field (Leinert et al. 1998).
We model the scattering by assuming a Henyey-Greenstein phase function 
(Henyey \& Greenstein 1941), $\Phi (\theta, g)$, for the 
angular distribution of scattered light.
The phase function is parameterised by the asymmetry parameter, 
$g \equiv g(\lambda)$, with $\lambda$ the wavelength.
We calculate $g$ by summing over grain types, and averaging, 
for each grain type $i$, the scattering phase function 
$g_i(\lambda, a)$ ($a = $ grain size), which we take from 
Laor \& Draine (1993), 
over our assumed grain size distribution, $n_i(a)$, weighted by the scattering 
cross-section, $\sigma^{scat} _i(\lambda, a)$.

We assume the dust follows the Galactic gas distribution, and use the gas 
model for neutral and molecular hydrogen described by Strong et al. (2000). 
Furthermore, the Galactic metallicity gradient is assumed to be 
exponential with radial scale length $\sim 4$ kpc.            

For the radiation field calculation, we assume a cylindrical geometry.
We simplify our calculations by adopting symmetry about the Galactic plane and 
in azimuth.
The maximum radial extent of the Galactic volume is taken to be 
$R_{max} = 20$ kpc, and the maximum height above the plane $z_{max}$ is set 
equal to 5 kpc.
We divide the total Galactic volume into elements $V_i$ of approximately 
equal size.
For each $V_i$, the radiation field, $u(\lambda)$, 
with full angular distribution, is calculated as we describe below.

The optical radiation field is obtained following the method of 
Kylafis \& Bahcall (1987). 
We first calculate the optical radiation field for absorption only (no 
scattering).
This radiation field is then used as input to calculate the amount
of light scattered only once.
The `once-scattered' radiation field is used as input to calculate 
the amount of
light scattered twice, and so forth for the desired number of scatterings 
in the calculation.
We have found absorption with the contribution by once- and twice-scattered
light is sufficient to adequately calculate the optical radiation field.
The total optical radiation field is obtained by summing these contributions.

The infra-red radiation field is obtained by using the total optical 
radiation field for each volume element to calculate the emissivity
for transient and equilibrium heating.
Then, for each $V_i$, we integrate over all volume elements to obtain the 
infra-red radiation field.
Subsequently, the infra-red radiation field for each volume element is used 
to calculate the re-absorbed infra-red emissivity, as the optical was used for 
the infra-red emissivity calculation. 
Another volume integration is again performed for each $V_i$ to obtain the
re-absorbed infra-red radiation field.
The total infra-red radiation field is the sum of these two components.

\begin{figure}[b]
\begin{minipage}[t]{0.48\textwidth}
\mbox{}\\
\centerline{\includegraphics[width=\textwidth]{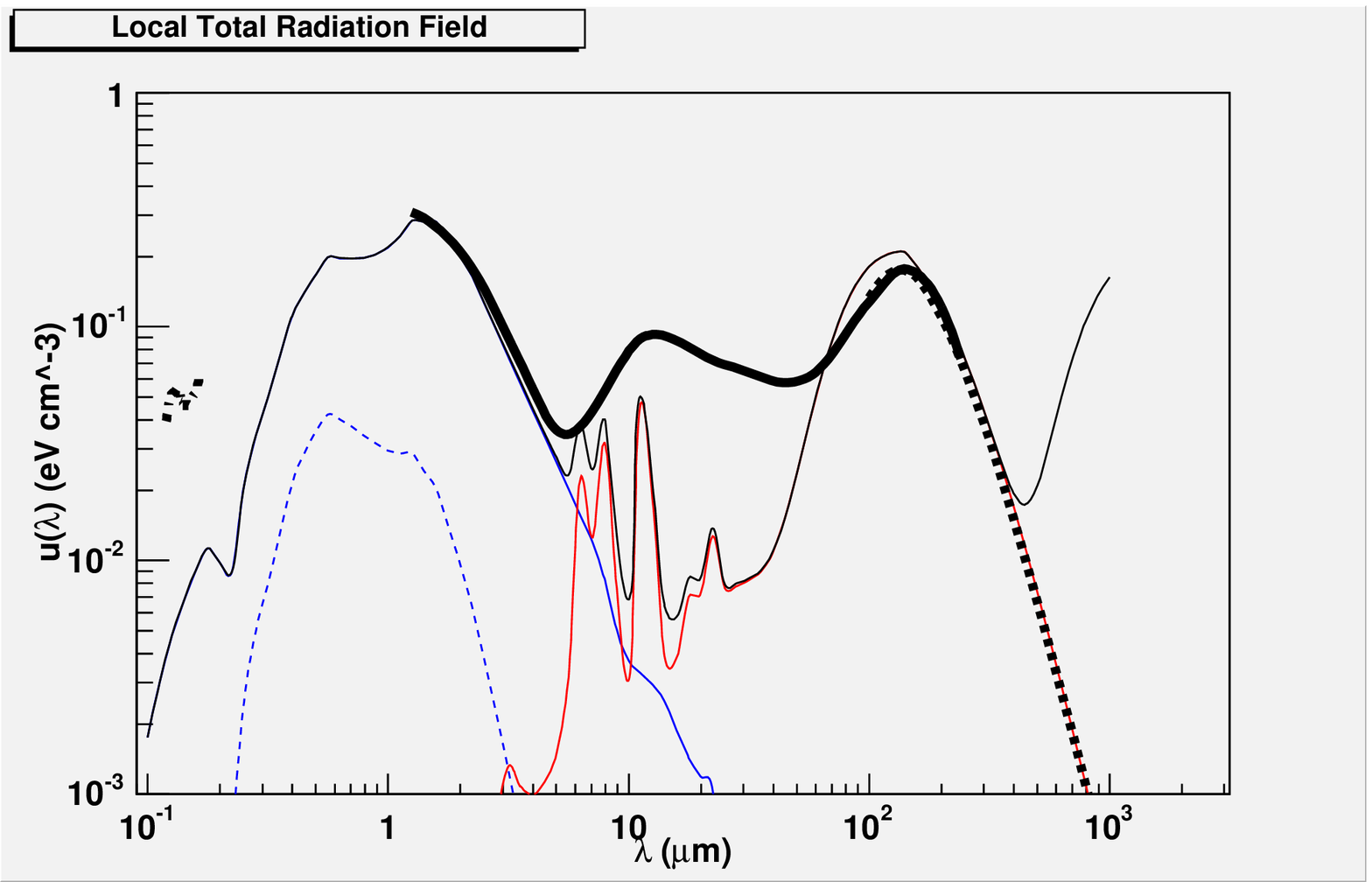}}
\caption{\label{fig:localISRF} 
Local radiation field modelled as described in the text.
Black line : total radiation field, including CMBR. 
Blue solid line : total optical. 
Blue dashed line : total scattered light. 
Red line : total infra-red. 
Data : thick dot-dashed line, Apollo; thick solid line, DIRBE; 
thick dashed line, FIRAS.}
\end{minipage}
\hfill
\begin{minipage}[t]{0.48\textwidth}
\mbox{}\\
\centerline{\includegraphics[width=\textwidth]{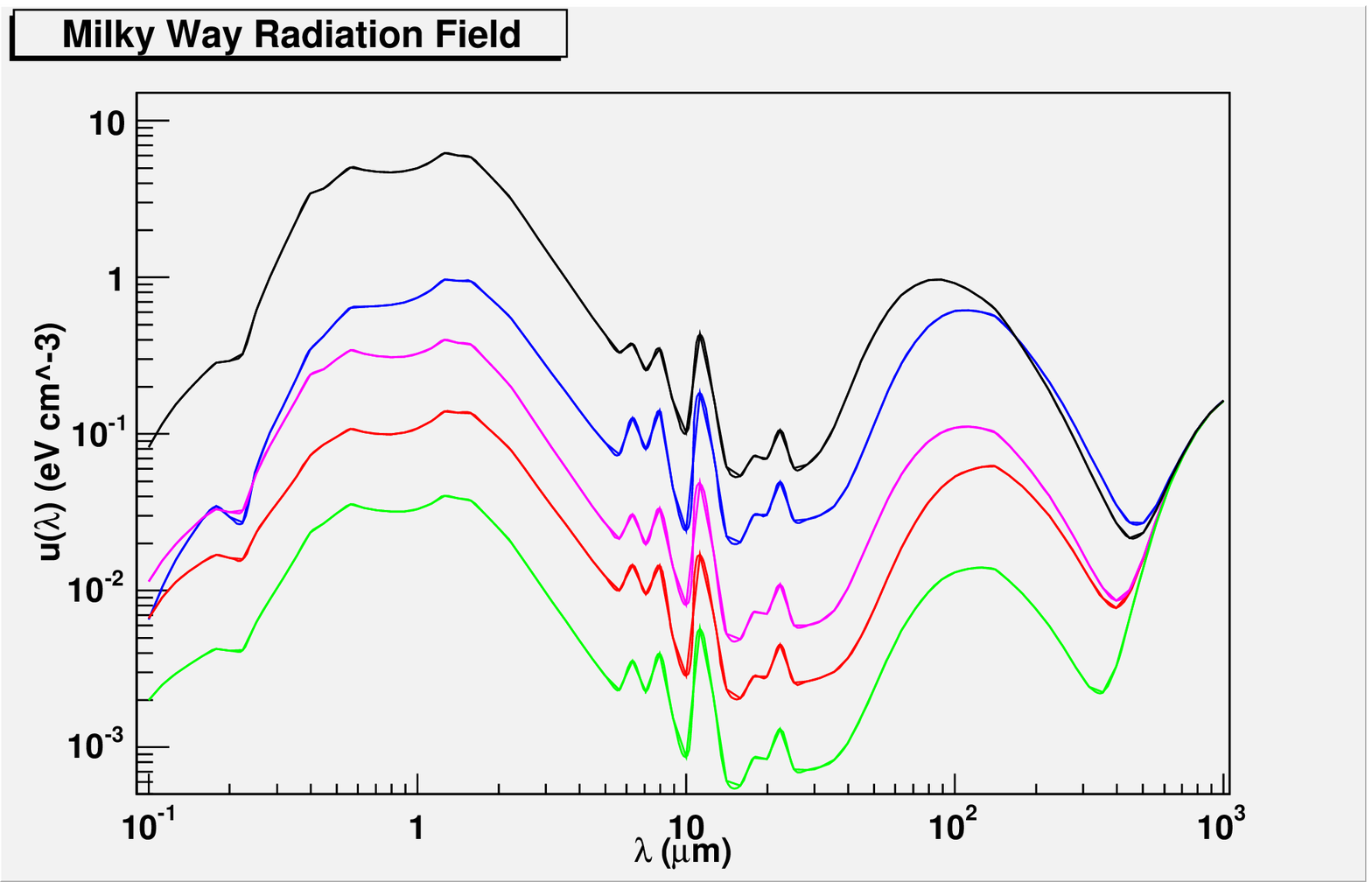}}
\caption{\label{fig:spatialISRF}Spatial variation of the total radiation field
throughout the Galaxy. 
Black line : total radiation field for GC. 
Magenta line : total radiation field for $R = 0$ kpc, $z = 5$ kpc. 
Blue line : total radiation field for $R = 4$ kpc, $z = 0$ kpc. 
Red line : total radiation field for $R = 12$ kpc, $z = 0$ kpc.
Green line : total radiation field for $R = 20$ kpc, $z = 0$ kpc.}
\end{minipage}
\end{figure}


Figure \ref{fig:localISRF} shows our calculated local radiation field, 
including the contribution by the cosmic microwave background (CMBR).
Also shown in the figure are observations by Apollo 
(Henry, Anderson, \& Fastie 1980), 
DIRBE (Arendt et al. 1998), and 
FIRAS (Finkbeiner et al. 1999).
The agreement with the observations is generally good. 
The scattered component of the optical radiation field is $\sim 10-20\%$ 
of the total optical radiation field between $\sim 0.2 - 2$ $\mu$m - in good
agreement with observations of the DGL 
(Leinert et al. 1998).
We note that our calculation falls significantly below the measured radiation
field in the ultraviolet regime, which may indicate a requirement to include
further source classes in the stellar distribution model.
Furthermore, the model underpredicts the infra-red from $\sim 10$ $\mu$m to 
$\sim 40$ $\mu$m.
However, the contamination by zodiacal light around $\sim 10$ $\mu$m in the 
DIRBE observations is uncertain, so this may not be such a problem.

Figure \ref{fig:spatialISRF} shows the ISRF at different locations in the 
Galaxy.
The radiation field is most intense in the inner Galaxy, and remains so even 
for distances above the Galactic plane of $\sim 5$ kpc, and probably more.
Interestingly, the emission longward of 100 $\mu$m for the in-plane $R = 0$ kpc 
and $R = 4$ kpc curves is about the same. 
This probably reflects the peak in the gas/dust distribution around 
$R \sim 4$ kpc.
However, further analysis will be required to ascertain if this is indeed the
true cause of this feature.

\section{Diffuse Gamma-Rays}
We will use our new determination for the Galactic ISRF to recalculate 
the contribution
of IC scattered gamma-rays to the total diffuse Galactic $\gamma$-rays,
using the GALPROP cosmic-ray propagation code (Strong et al. 2000, 2004).
Results will be presented at this conference.

When completed the new ISRF will be made publicly available as part of the
GALPROP package (see http://www.mpe.mpg.de/$\sim$aws/aws.html).

%


\begin{thebibliography}{99}

\bibitem{allen2000} Allen's Astrophysical Quantities (4th edition), ed. Cox, A. N., AIP Press, Springer-Verlag (2000). 

\bibitem{arendt1998} Arendt, R. G., et al., ApJ 508, 74 (1998). 

\bibitem{cohen1993} Cohen, M., AJ 105, 1860 (1993).

\bibitem{cohen1994} Cohen, M., AJ 107, 582 (1994).

\bibitem{cohen1995} Cohen, M., ApJ 444, 874 (1995).

\bibitem{draine2001} Draine, B. T. \& Li, A., ApJ 551, 807 (2001).

\bibitem{finkbeiner1999} Finkbeiner, D., Davis, M. \& Schlegel, D. J., 
ApJ 524, 867 (1999).

\bibitem{freudenreich1998} Freudenreich, H. T., ApJ 492, 495 (1998).

\bibitem{girardi2002} Girardi, L., et al., A\&A 391, 195 (2002).

\bibitem{henry1980} 
Henry, R. C., Anderson, R. C. \& Fastie, W. G., ApJ 239, 859 (1980).

\bibitem{henyey1941} Henyey, L. G. \& Greenstein, J. L., ApJ 93, 70 (1941).

\bibitem{kylafis1987} Kylafis, N. D. \& Bahcall, J. N., ApJ 317, 637 (1987).

\bibitem{laor1993} Laor, A. \& Draine, B. T., ApJ 402, 441 (1993).

\bibitem{leinert1998} Leinert, Ch., et al., A\&AS 127, 1 (1998).

\bibitem{li2001} Li, A. \& Draine, B. T., ApJ 554, 778 (2001).
 
\bibitem{lopez2001} L\'{o}pez-Corredoira, M., et al., A\&A 373, 139 (2001).

\bibitem{lopez2002} L\'{o}pez-Corredoira, M., et al., A\&A 394, 883 (2002).

\bibitem{lopez2004} L\'{o}pez-Corredoira, M., et al., A\&A 421, 953L (2004).



\bibitem{ojha2001} Ojha, D. K., MNRAS 322, 426 (2001).

\bibitem{strong2000} Strong, A. W., Moskalenko, I. V. \& Reimer, O., 
ApJ 537, 763 (2000).

\bibitem{strong2004} Strong, A. W., Moskalenko, I. V. \& Reimer, O., 
ApJ 613, 962 (2004).

\bibitem{wainscoat1992} Wainscoat, R. J., et al., ApJS 83, 111 (1992).

\bibitem{weingartner2001} Weingartner, J. C. \& Draine, B. T., ApJ 548, 296 
(2001).

\bibitem{zubko2004} Zubko, V., Dwek, E. \& Arendt, R. G., ApJS 152, 211 (2004).

\end{thebibliography}
\end{document}